# Citation of scientific evidence from video description and its association with attention and impact

Pablo Dorta-González [a,*] and María Isabel Dorta-González [b]

[a] Institute of Tourism and Sustainable Economic Development (TIDES), University of Las Palmas de Gran Canaria, Las Palmas de Gran Canaria, Spain

[b] Departamento de Ingeniería Informática y de Sistemas, University of La Laguna, La Laguna, Spain

* Correspondence: University of Las Palmas de Gran Canaria. Campus de Tafira, 35017 Las Palmas de Gran Canaria, Spain. pablo.dorta@ulpgc.es

This study investigates how YouTube content creators utilize scientific evidence in videos. Log-linear regression examines the influence of alternative communication channels on video creators in Biotechnology, using data from 81,302 papers (2018-2023). This reveals a positive association with news articles and Wikipedia pages, but a negative association with scientific papers, policy documents, and patents. Despite the potential for enriching discussions, science video creators seem to favor materials with wider public attention over influential science, technology, and policy papers. These findings suggest a need for improved dissemination strategies for scientific research. Authors, universities, and journals should consider how their work can be made more accessible and engaging for science communicators on video.

# Introduction

In addition to its scientific impact, which is commonly gauged by citation metrics, scientific research is valuable in terms of its wider societal impact. The impact of research on all facets of society is referred to as societal impact. The societal impact of research was described as its influence on education, society, culture, or the economy in a study by Wilsdon et al. (2015). As the NISO Alternative Assessment Metrics (NISO 2016) point out, altmetrics offer a quantitative way to gauge the wider impact of publications.

The assessment of the societal impact of scholarly research has changed significantly with the development of digital scholarly communication. This has led to a more varied strategy that includes a greater variety of academic publications and innovative communication techniques (see Bornmann 2013; de Rijcke et al. 2016; Bornmann & Haunschild 2019). The UK adopted the Research Excellence Framework to evaluate the quality of research being carried out at universities (REF 2021). The assessment of influence outside the scientific domain is given considerable weight in this paradigm, accounting for 25% of the overall evaluation. According to Khazragui & Hudson (2015), this involves assessing how research affects public policy, services, the economy, society, culture, health, environment, and general quality of life.

The necessity to reevaluate altmetrics in connection to effect is becoming more widely acknowledged by the research community (Spaapen & van Drooge 2011; Joly et al. 2015; Morton 2015). Recent research indicates that altmetrics are better understood as instruments for examining how research interacts with society and how knowledge flows beyond academic borders rather than serving as direct markers of impact (Haustein et al. 2016; Ravenscroft et al. 2017). Scholars who have called for a reconsideration of altmetrics (Robinson-García et al. 2018; Wouters et al. 2019) have further supported this viewpoint, and research is still being done to improve this developing framework (Costas et al. 2021; Alperin et al. 2023; González-Betancor & Dorta-González 2023).

The impact of research on society is evident in the incorporation of scientific papers into policy documents (Yu et al. 2023) and patent documents (Oldham 2022). Citing research in policy documents also adds legitimacy and offers insights into academic research and policymaking (Bornmann et al. 2016). Citing research publications in patent documents emphasizes how academic research influences innovation. Oldham (2022) draws attention to the connection between innovation and research, stressing the significance of this relationship in exposing how scientific knowledge is incorporated into patents.

The impact of research on society is evident when scientific papers are mentioned on social networks (Dorta-González 2023). Although researchers have examined the motivations behind research mentions on social media platforms such as X, research on the factors influencing research citations on video channels is lacking. According to Shaikh et al. (2023), YouTube creators, particularly those with academic backgrounds, often utilize scientific papers to support their debates, lending legitimacy and providing insightful information. A dedication to openness and intellectual integrity is demonstrated by the references included in the descriptive metadata of their videos, which enables viewers to access the sources and delve deeper into the subjects covered (Welbourne & Grant 2016). In addition to improving the educational value of their videos, content makers who engage with academic literature in this way also help their communities spread critical thinking and accurate information (Amarasekara & Grant 2019; Shaikh et al. 2023).

According to some authors, research publications can become more visible and receive more citations if they use video as a technique for information transmission and scholarly dissemination (Shaikh et al. 2023). Furthermore, video abstracts are seen to be more enticing to readers, which helps with comprehension and is associated with greater rates of article reads (Bonnevie et al. 2023). Nonetheless, it's possible that videos don't just influence citations, but rather that papers with more citations are chosen for video production, or that the two factors have an impact on one another (Kohler & Dietrich 2021).

This research adopts a method by examining how the creation of science videos is influenced by the dissemination of science through alternative communication channels, whereas previous studies have mostly looked at the effect of videos on citations. We use log-linear regression to examine the possible effects of various communication channels on YouTube citations to accomplish this objective.

**Literature review**

*YouTube and the spread of scientific information*

Due to their substantial influence on audience engagement and the spread of scientific knowledge, YouTube and online science communication have garnered a lot of attention lately (Hutchinson 2017; Iqbal 2024). With an enormous user base and high daily activity (O'Neil-Hart & Blumenstein 2016), YouTube has quickly grown to become one of the largest websites in the world (Sui et al. 2022). Notably, YouTube is widely used by people of all ages, but especially by those between the ages of 18 and 49, where eight out of ten people watch videos monthly (GMI Blogger 2024).

Science and technology are major categories in the YouTube content landscape, accounting for around 4% of all video uploads and ranking seventh overall (Hutchinson 2017; Amarasekara & Grant 2019). Although YouTube's large subscriber base and viewership analytics demonstrate the appeal of science-themed channels (Agarwal 2024), little is known about how prevalent and active scientists are on the platform. Mixed results have come from attempts to investigate how scientists use YouTube. Although some surveys show that scientists use media sharing sites like YouTube and Flickr seldom (Collins et al. 2016), others reveal a more significant presence, with almost half of the scientists polled using these sites at least once a week (Nikiphorou et al. 2017). Furthermore, there is a noticeable gap in the literature due to the paucity of studies on the reasons behind and trends in the distribution of science-related information on YouTube (SCIMEP 2016; Amarasekara & Grant 2019).

*User engagement with science content on YouTube*

Scholars seeking to comprehend the processes driving audience interaction have paid close attention to how viewers interact with online science videos, especially those on YouTube. Yang et al. (2022) looked at a number of variables that affect how users interact with science videos on the internet. Their research highlighted the importance of elements, including presentation style, relevancy, and content quality, in drawing in and retaining visitors. Welbourne & Grant (2016) have looked at the factors influencing the popularity of science communication channels and videos on YouTube. Their findings showed how important factors like video length, production quality, and presenter talent are in holding viewers' attention.

In a related study, Velho et al. (2020) examined the association between a number of variables and the number of people who view science-related YouTube videos. According to their research, factors like video length, header image, and title clarity all have an impact on how many people watch a video. The profiles, difficulties, and motives of science YouTubers were also examined by Velho & Barata (2020), who provided insight into the many traits and goals of content producers in this field.

Huang & Grant (2020) concentrated on the elements of storytelling that make science videos on YouTube so successful. According to their analysis, the main elements influencing how audiences interact with science content are narrative structure, emotional appeal, and visual storytelling. Additionally, to clarify the complex mechanisms behind viewers' perceptions and reactions to science videos on YouTube, Dubovi & Tabak (2021) investigated the interaction between emotional and cognitive involvement.

*Video channel citations and impact on science dissemination*

Citing research articles in YouTube videos can increase scientific impact and visibility. Citations to research articles are frequently included in the descriptions of videos on websites like YouTube, which can be very helpful in sharing research findings with a wider audience (Shaikh et al. 2023). According to Welbourne & Grant (2016), adding citations to video content can further enhance the videos' legitimacy and educational value, which may boost viewers' trust.

The fact that research publications from disciplines like psychology and cognitive sciences, biological sciences, and medical and health sciences are among the most often cited in YouTube videos suggests that the public is more interested in and involved with these subjects (Shaikh et al. 2023). Understanding the several content characteristics that affect the popularity of science communication videos on YouTube can assist in increasing the popularity of the videos and, in turn, the citation of the papers they feature (Welbourne & Grant 2016). Furthermore, because videos facilitate the quick online sharing of study findings, using them as a dissemination medium for studies can reduce the time it takes to put evidence into practice. When videos are made to be interesting and include emotional content that promotes sharing, this can work especially well (Kiriya 2016).

According to Kousha et al. (2012), Yang et al. (2022), and Shaikh et al. (2023), citing research papers in video channels can help spread scientific findings in several ways, including: increasing research papers' visibility and reaching a wider and more diverse audience; improving public knowledge and engagement with science; encouraging discussion and feedback between viewers and researchers; igniting viewers' curiosity and interest in science; and encouraging cooperation and creativity between researchers and video creators.

By examining citation patterns across fields, Dorta-González (2025) investigates how YouTube content producers incorporate scientific information into their videos. The study uses Principal Component Analysis to characterize the two primary components of evidence citation used by different stakeholders in science communication. While the second component concentrates on a paper's social effect and impact, which are defined by the paper's quality and scientific significance, the first component increases a paper's visibility by attracting social attention.

**Methodology**

*Objective*

No previous studies have examined the association between citations of research papers in video communication channels and their citations in research journals, policy documents, patent documents, and other altmetrics sources. The aim of this study is therefore to describe this association.

*Data*

Information was extracted from Altmetric.com's document search interface. Altmetric.com uses a content tracking method that concentrates on well-vetted YouTube channels as part of its data collection procedure. These channels were picked because they are pertinent to the dissemination of research and intellectual output. The YouTube collector looks for direct links to academic outputs in the description area of videos whenever they are uploaded to one of the monitored channels.

The Field-of-Research (FoR) code scheme was employed to guarantee relevance to particular topics that are known to have regular video communication. Because the field classification system is AI-driven and its accuracy is unknown, it is crucial to recognize that it is not flawless. The discipline with the most citations in videos is Medical Biotechnology, represented by the code 3206.

To capture current trends while preserving historical context, the focus was on research publications published between 2018 and 2023. This was especially important for citations that usually have a longer gestation time, such as those found in Wikipedia, patents, and policy documents. To ensure that only scientific publications that significantly advance their respective fields are included, the search was limited to "research articles". As a result, 81,302 papers were considered in the final dataset.

*Methods*

Table 1 describes the quantitative variables. Because of the features of the data, this study uses Spearman correlation instead of Pearson correlation in terms of statistical methods. Spearman correlation is a non-parametric measure used to assess the strength and direction of monotonic correlations between two variables. Since Spearman's correlation does not depend on linear correlations or a normal distribution, it can be applied to variables that may not satisfy the assumptions of Pearson's correlation. As a result, Spearman's correlation provides a robust indication of association that is not influenced by these assumptions and is better suited to the goals of this investigation.

[PLACE TABLE 1 HERE]

Although the number of citations a paper obtains in videos was thought to be modeled by a Poisson distribution, the data revealed overdispersion (variance above the mean). After adding one to each value, the analysis changed to a Gamma distribution, which was then applied to the data. Dealing with positively skewed continuous data, like the recollection of a rare occurrence, this model works especially well. Gamma regression uses a maximum likelihood method for parameter estimation.

Thus, the model is written as

$$\log(Y_i) = \beta_0 + \beta_1 x_{i1} + \ldots + \beta_j x_{ij},$$

which is equivalent to

$$Y_i = e^{\beta_0 + \beta_1 x_{i1} + \ldots + \beta_j x_{ij}} = e^{\beta_0} e^{\beta_1 x_{i1}} \ldots e^{\beta_j x_{ij}}.$$

Interpretation of these parameters is as follows. The $e^{\beta_j}$ is the multiplicative effect on the mean of Y when $x_j$ increases by 1 unit. Therefore:

(1) If $\beta_j = 0$, then $e^{\beta_j} = 1$, and the expected count, Y, and $x_j$ are not related.

(2) If $\beta_j > 0$, then $e^{\beta_j} > 1$, and the expected count is $e^{\beta_j}$ times larger than when $x_j = 0$.

(3) If $\beta_j < 0$, then $e^{\beta_j} < 1$, and the expected count is $e^{\beta_j}$ times smaller than when $x_j = 0$.

**Results**

This section presents the results from a Log-linear Regression Model for all research articles in Biotechnology in the cohort 2018-2023 (N=81,302). Several statistics regarding the variables taken into consideration are displayed in Tables 2 and 3. According to the paper's acknowledgments section, 63% of the research publications were funded in some way (Table 2).

[PLACE TABLES 2 AND 3 HERE]

The data distributions are extremely skewed towards zero (see Table 3). Bibliometric data frequently exhibit this skewness, which calls for the application of particular statistical methods.

As previously mentioned, when the assumptions of ordinary least squares are broken, Gamma regression provides a reliable substitute for modeling continuous data with positive skewness. Furthermore, Spearman correlations (see Table 4) are especially well-suited for evaluating the associations between variables since they are less susceptible to the effects of extreme values and non-normality. As can be observed in Table 4, the number of readers on Mendeley strongly correlates with the number of citations in Dimensions.

[PLACE TABLE 4 HERE]

The goodness of fit for the log-linear regression model using a Gamma distribution is presented in Table 5. The null model, which includes only a constant and no explanatory variables, serves as a reference. To compare the deviance between the null model and the adjusted model, a $G^2$ test for goodness of fit, corresponding to -2·log-likelihood, was conducted. A smaller $G^2$ value is preferred. The $G^2$ test yielded a very small p-value (< 0.0001), indicating that the full model significantly outperforms the null model. Moreover, the R² (McFadden) is 0.49. This is a coefficient between 0 and 1 that measures how well the model is adjusted. This coefficient is equal to 1 minus the ratio of the likelihood of the adjusted model to the likelihood of the null model.

[PLACE TABLE 5 HERE]

The equation of the model in Table 6 is the following:

Pred (YouTube + 1) = exp (1 + 0.005 · News - 0.004 · Blog - 0.019 · Policy - 0.008 · Patent + 0.000 · X - 0.031 · Facebook + 0.013 · Wikipedia - 0.001 · Citations - 0.327 · Unfunded - 0.805 · Year_2019 - 0.822 · Year_2020 - 0.836 · Year_2021 - 0.842 · Year_2022)

[PLACE TABLE 6 HERE]

For each extra unit in the explanatory variable, the coefficient interpretations shown in Table 6 indicate the percentage change in the number of citations inside video channels. Similarly, when the explanatory variable increases by one unit, the incidence ratio in the last column ($e^{coef}$) indicates the multiplicative effect on the average number of citations in YouTube videos.

With p-values less than 0.01, every variable in the model shows strong significance, except for the quantity of citations in blogs. The lack of funding considerably reduces the number of citing videos. Citations in YouTube videos are 33% lower for studies without financial sources than for those that have (incidence ratio of 0.72).

With incidence ratios of 1.013 and 1.005, respectively, Wikipedia and News rank top among the quantitative factors that have a positive impact on citations in videos. As a result, video citations rose by 1.3% for every Wikipedia reference and by 0.5% for every news reference. Conversely, with incidence ratios ranging from 0.970 to 0.999, Facebook mentions, citations in policy papers, patents, and research journals are the quantitative variables that have the biggest negative impact on citations in videos.

For instance, every policy citation reduces video references by 2%, and every Facebook posting reduces video citations by 3%. Similarly, every patent reference reduces video mentions by less than 1%, and every scientific citation reduces video citations by only 0.1%. Despite their small size, these consequences are significant when all other factors are held constant. Interestingly, given how often it is cited in scholarly works compared to patents or policies, this conclusion is by no means inconsequential. Accordingly, a 10% decrease in the number of citing videos is associated with 100 citations in research publications. This confirms the previous findings that social attention has a greater influence on video producers than scientific or technological advancements.

**Discussion**

YouTube content creators, particularly those with research or scholarly backgrounds, frequently utilize scientific studies to substantiate their claims and enhance discussions of concepts, hypotheses, discoveries, or practical applications. Biotechnologists frequently reference scientific articles in their video descriptions and are especially dependent on published research (Shaikh et al. 2023). This paper aimed to analyze, for this discipline, the use of scientific evidence in video communication and the influence of scientific research on content creators on YouTube.

There is some evidence that altmetric scores do not correlate with scientific citations (Spaapen & van Drooge 2011; Joly et al. 2015; Morton 2015), but this is not consistent with other societal impacts. Furthermore, no previous studies have examined the association between altmetric scores and citations in video channels. This paper described this association using research articles published between 2018 and 2023 in the discipline with the highest prevalence of citations in videos, Biotechnology, including both those with citations in videos and those without.

We found a positive association with News. The results of the regression show that an increase in the number of news items by ten was associated with a 5% increase in the number of citations in YouTube videos. We also found a positive association with Wikipedia. The regression findings indicate that increasing citations in Wikipedia by one was associated with a 1.3% increase in citations in YouTube videos. However, a negative association was found with the number of citations in scientific papers, policy documents, and patents.

The observed association between the citation frequency of research papers on Wikipedia and in YouTube videos can be explained. Given that content creators often use both Wikipedia and YouTube as sources of information, a paper cited on Wikipedia is likely to be encountered by YouTube creators who may incorporate it into their videos, highlighting the interconnected nature of these platforms as sources of information.

The positive association between the number of news items about a scientific paper and the number of YouTube videos citing that paper could be attributed to the interconnectedness of traditional media and video communication, where news coverage can act as a catalyst for increased visibility of scientific research (Shaikh et al. 2023). News articles often serve as accessible summaries of scientific research, making the findings more widely known and increasing the likelihood that they will be cited in various media, including YouTube videos. Furthermore, news coverage may increase the visibility and credibility of scientific papers, leading to a higher likelihood of their citation in various online platforms, including YouTube, where content creators may seek authoritative sources for their videos.

Conversely, the negative association observed between the number of citations a paper receives in other papers, policy documents, patents, and the number of YouTube videos citing that paper highlights the differences in audience preferences, content requirements, credibility considerations, and information accessibility between academic literature, policy documents, patents, and YouTube videos that influence the likelihood of citations in YouTube videos (Sui et al. 2022). Papers cited in academic literature, policy documents, and patents are typically aimed at specialized audiences such as researchers, policy makers, and professionals in specific fields. These documents prioritize technical rigor, methodology, and detailed information relevant to their respective fields. In contrast, YouTube videos are aimed at a broader audience seeking simplified explanations and entertainment (Yang et al. 2022). Therefore, papers cited in academic literature, policy documents, and patents may not match the content preferences or goals of YouTube creators and their audiences, leading to fewer citations in YouTube videos.

The apparent contradictory results regarding the associations between citations in different social networks and the number of YouTube videos citing a paper highlight the complexity of social media dynamics, audience engagement patterns, and content preferences. Content shared on Facebook may not be consistent with the type of content typically found in YouTube videos, leading to a negative association. Conversely, the lack of association between citations on X and YouTube videos may be because the X platform, characterized by short-form and real-time communication, may not be conducive to in-depth discussions or the dissemination of content that would lead to citations in YouTube videos. On the other hand, the audiences of Facebook, X, and YouTube may have different demographics and interests.

We found that papers lacking funding sources receive 33% fewer citations in YouTube videos compared to those acknowledging funding. This significant decrease in citations for unfunded papers could be attributed to the limitations in resources and opportunities available to researchers without financial support (Dorta-González & Dorta-González 2023), ultimately affecting the dissemination and impact of their work in the online video community.

There are some limitations to this study. The scientific disciplines range significantly in terms of public exposure, communication styles, and citation conventions. These disciplinary particularities make it difficult to extrapolate the findings to other domains not covered by the study, especially those with significantly different dissemination dynamics or scholarly prominence. Furthermore, even if YouTube is currently the most popular and accessible video platform worldwide, scientific communication in audiovisual format occurs through other channels as well. Scientific material can also be found on platforms like Vimeo, Instagram Reels, and TikTok, which may have distinct interaction dynamics or citation patterns. As a result, it is important to interpret the results cautiously and within the context of the platforms and discipline under consideration.

## Conclusions

Some authors claim that using videos as a technique for knowledge transfer and scholarly dissemination can raise research publications' visibility and, consequently, their citation counts (Shaikh et al. 2023). Furthermore, videos are more visually appealing to readers, make it easier for them to comprehend the substance of the research, and are linked to higher article reading rates (Bonnevie et al., 2023). Citations in videos and citations in other sources, however, may have a reciprocal or inverse relationship; that is, videos may not create citations, but rather papers with more citations may be selected to create videos, or the two may reinforce one another (Kohler & Dietrich 2021). For this reason, this study examined the opposite link, that is, how the dissemination of science through other communication channels affects the creation of videos.

This study has found a small but significant and positive association between the citations a paper receives in news items and Wikipedia and its citations in YouTube videos. On the other hand, there is a weak but negative correlation between citations in videos and citations in other scholarly articles, policy documents, and patents. Similarly, there is a weak but negative correlation between the number of citations in videos and the frequency of social media conversations.

These findings imply that the prominence of scientific publications in YouTube videos may benefit from their distribution via non-scientific channels like news sites and Wikipedia. Additionally, there may be a gap between the scientific community and the audiences of these platforms, as evidenced by the negative link between citations in videos, their citation rates in scientific publications, policy documents, and patents, and debates on social media. This emphasizes how crucial it is to take into account the various methods of sharing scientific knowledge as well as any potential trade-offs involved.

The study's main conclusion is that, although creators of scientific YouTube content could add to discussions by citing important academic works, they usually opt to cite more readily available content, such as news articles and Wikipedia entries. This suggests that there may be a discrepancy between the ease of access to scientific knowledge for YouTube creators and their audience and the way in which scientific knowledge is distributed.

As a recommendation, this study highlights the need to bridge the gap between impactful scientific research and its accessibility for YouTube creators. To achieve this, stakeholders, including authors, universities, and journals, should re-evaluate dissemination strategies. The focus should shift towards making research more accessible and engaging for science communicators, ultimately fostering a more informed and enriched discussion on YouTube.

Table 1. Description of quantitative variables: Types of references to research papers

| Type | Measure | Variable | Description | Source |
|---|---|---|---|---|
| Societal Influence | Scientific impact | Citations | Number of citations in scholarly publications | Dimensions[1] |
| | Technological impact | Patent | Number of citations in patent documents | Altmetric |
| | Impact on policies | Policy | Number of citations in policy documents | Altmetric |
| | Impact on Readers | Readers | Number of times a paper has been saved in bibliographic reference managers | Mendeley[1] |
| Societal Attention | Media attention | News | Number of times a paper has been cited in digital newspapers | Altmetric |
| | Blogging attention | Blog | Number of times a paper has been cited in blogs | Altmetric |
| | Social media attention | X, Facebook | Number of conversations in social networks | Altmetric |
| | Encyclopedic attention | Wikipedia | Number of references in pages of Wikipedia | Wikipedia[1] |
| | Video attention | YouTube | Number of times a paper has been cited in video channels of YouTube | YouTube[1] |

Note[1]: Altmetric was used as an indirect or secondary source of information.

Table 2. Statistics for the qualitative variables in the regression model

| Variable | Category | Frequency | % |
|---|---|---|---|
| Funding | Funded | 51,472 | 63.3 |
| | Unfunded | 29,830 | 36.7 |
| Year | 2018 | 11,669 | 14.4 |
| | 2019 | 11,599 | 14.3 |
| | 2020 | 13,294 | 16.4 |
| | 2021 | 15,643 | 19.2 |
| | 2022 | 15,718 | 19.3 |
| | 2023 | 13,379 | 16.5 |

Table 3. Statistics for research articles in Biotechnology in 2018-2023 (N=81,302)

| Variable | Obs. | Missing | min | max | Mean | SD |
| --- | --- | --- | --- | --- | --- | --- |
| YouTube + 1 | 81,302 | 0 | 1 | 142 | 1.020 | 0.702 |
| News | 81,302 | 0 | 0 | 548 | 0.728 | 6.965 |
| Blog | 81,302 | 0 | 0 | 43 | 0.093 | 0.625 |
| Policy | 81,302 | 0 | 0 | 33 | 0.016 | 0.244 |
| Patent | 81,302 | 0 | 0 | 223 | 0.146 | 1.528 |
| X | 81,302 | 0 | 0 | 29,750 | 8.223 | 167.453 |
| Facebook | 81,302 | 0 | 0 | 25 | 0.139 | 0.706 |
| Wikipedia | 81,302 | 0 | 0 | 23 | 0.039 | 0.377 |
| Readers | 81,302 | 0 | 0 | 4,099 | 36.896 | 75.341 |
| Citations | 81,302 | 0 | 0 | 3,836 | 21.723 | 49.575 |

Table 4. Spearman correlation coefficient for the quantitative variables in the regression model (N=81,302)

|  | News | Blog | Policy | Patent | X | Facebook | Wikipedia | Readers | Citations |
| --- | --- | --- | --- | --- | --- | --- | --- | --- | --- |
| News | 1 | 0.687 | 0.150 | 0.053 | 0.184 | 0.271 | 0.202 | 0.177 | 0.143 |
| Blog | 0.687 | 1 | 0.122 | 0.076 | 0.209 | 0.337 | 0.210 | 0.233 | 0.194 |
| Policy | 0.150 | 0.122 | 1 | 0.008 | 0.070 | 0.084 | 0.083 | 0.114 | 0.101 |
| Patent | 0.053 | 0.076 | 0.008 | 1 | 0.019 | 0.058 | 0.075 | 0.185 | 0.179 |
| X | 0.184 | 0.209 | 0.070 | 0.019 | 1 | 0.158 | 0.061 | 0.078 | 0.060 |
| Facebook | 0.271 | 0.337 | 0.084 | 0.058 | 0.158 | 1 | 0.107 | 0.199 | 0.170 |
| Wikipedia | 0.202 | 0.210 | 0.083 | 0.075 | 0.061 | 0.107 | 1 | 0.187 | 0.156 |
| Readers | 0.177 | 0.233 | 0.114 | 0.185 | 0.078 | 0.199 | 0.187 | 1 | 0.875 |
| Citations | 0.143 | 0.194 | 0.101 | 0.179 | 0.060 | 0.170 | 0.156 | 0.875 | 1 |

Table 5. Goodness of fit for the log-linear regression model using a Gamma distribution

| Statistics | Null model | Adjusted model |
|---|---|---|
| Obs. | 81,302 | 81,302 |
| DF | 81,301 | 81,289 |
| $G^2$ (-2·log-lik) | 216,177 | 109,752 |
| p-value (Pr > Chi²) |  | < 0.0001 |
| R² (McFadden) | 0.000 | 0.492 |
| AIC | 216,181 | 109,782 |

Table 6. Model parameters for the log-linear regression model using a Gamma distribution

| Source | Coef. | SE | Wald Chi$^2$ | Pr > Chi² | 95% CI | Incidence ratio ($e^{coef}$) |
|---|---|---|---|---|---|---|
| Intercept | 1.000 |  |  |  |  |  |
| News | 0.005 | 0.000 | 169.285 | < 0.0001 | 0.004 to 0.005 | 1.005 |
| Blog | -0.004 | 0.004 | 1.400 | 0.237 | -0.012 to 0.003 | 0.996 |
| Policy | -0.019 | 0.006 | 10.453 | 0.001 | -0.030 to -0.007 | 0.981 |
| Patent | -0.008 | 0.001 | 112.237 | < 0.0001 | -0.009 to -0.006 | 0.992 |
| X | 0.000 | 0.000 | 7.682 | 0.006 | 0.000 to 0.000 | 1.000 |
| Facebook | -0.031 | 0.002 | 189.664 | < 0.0001 | -0.035 to -0.026 | 0.970 |
| Wikipedia | 0.013 | 0.005 | 7.987 | 0.005 | 0.004 to 0.022 | 1.013 |
| Readers | 0.000 | 0.000 |  |  |  |  |
| Citations | -0.001 | 0.000 | 1539.433 | < 0.0001 | -0.001 to -0.001 | 0.999 |
| Funded | 0.000 | 0.000 |  |  |  |  |
| Unfunded | -0.327 | 0.003 | 9648.602 | < 0.0001 | -0.333 to -0.320 | 0.721 |
| Year_2018 | 0.000 | 0.000 |  |  |  |  |
| Year_2019 | -0.805 | 0.005 | 31322.401 | < 0.0001 | -0.814 to -0.796 | 0.447 |
| Year_2020 | -0.822 | 0.004 | 37615.864 | < 0.0001 | -0.830 to -0.814 | 0.439 |
| Year_2021 | -0.836 | 0.004 | 44567.406 | < 0.0001 | -0.844 to -0.829 | 0.433 |
| Year_2022 | -0.842 | 0.004 | 45361.711 | < 0.0001 | -0.850 to -0.834 | 0.431 |
| Year_2023 | 0.000 | 0.000 |  |  |  |  |
| Scale | 0.000 | 0.000 | 0.000 | < 0.0001 | 0.000 to 0.000 | 1.000 |